\RequirePackage{fix-cm}
\documentclass[twocolumn,epjc3]{svjour3}  
\smartqed  
\RequirePackage{graphicx}
\journalname{Eur. Phys. J. C}
\begin{document}

\title{The mass spectra and decay properties of dimesonic states, using the Hellmann potential
}


\author{Ajay Kumar Rai
        \and
        D. P. Rathaud 
}

\thankstext{e1}{e-mail: raiajayk@gmail.com}


\institute{Department of Applied Physics, Sardar Vallabhbhai National Institute of Technology, Surat, Gujarat, India -395007 \label{addr1} 
}

\date{Received: date / Accepted: date}

\maketitle

\begin{abstract} 
Mass spectra of the dimesonic (meson -
antimeson) molecular states are computed using the Hellmann potential in variational approach, which consists of relativistic correction to kinetic energy term as well as to the potential energy term. For the study of molecular bound state system, the Hellmann potential of the form  $V(r)=-\frac{\alpha_{s}}{r} + \frac{B e^{-Cr}}{r}$  is being used.  The one pion exchange potential (OPEP) is also incorporated in the mass calculation. The digamma decay width and decay width of the dimesonic system are evaluated using the wave function. The experimental states such as $f_{0}(980)$, $b_{1}(1235)$, $h_{1}(1380)$, $a_{0}(1450)$, $f_{0}(1500)$, $f_{2}'(1525)$,
$f_{2}(15 \\
65)$, $h_{1}(1595)$, $a_{2}(1700)$, $f_{0}(1710)$, $f_{2}(1810)$ are compared with dimesonic states.  Many of these states (mass-
es and their decay properties) are close to our theoretical predictions.

\end{abstract}

\section{Introduction}
\label{intro}
The spectroscopy of hadrons imparts information of basic force of nature. The theory of Quantum Chromo Dynamics (QCD) is a tool to understand the strong force in which hadron spectroscopy plays a key role in non-perturbative and perturbative regime. The new experimental developments at Belle, BES, CLEO, CDF, LHC, BABAR brought out enormous data and came with large numbers of surprises \cite{21,1,2,3,4}.  Number of new states are observed which do not fit in the conventional $q\overline{q}$  scheme,  mainly in meson sector. All these exotic states which do not fit in qqq and $q\overline{q}$  scheme require extra theoretical attention \cite{Brambilla-2,Yong-Liang Ma,5,R.L. Jaffe,6}. \\
Recently, in partial wave analysis (PWA) of $ J/\psi\rightarrow\gamma\eta\eta $; BES III collaboration observed most promising candidates for glueballs or multiquark structure below $2.5$ GeV \cite{Ablikim,Hai-Bo Li,Liaoyuan Dong}.  In $\pi^{-}P\rightarrow\eta\pi^{0}\eta $  channel GAMS group at CERN reported state near 1400 MeV \cite{D. Alde,Daniel S. Carman}. BNLE-
852  collaboration reported another state near $1600$ MeV in $\pi^{-}p$ reaction having promising hybrid structure \cite{Daniel S. Carman,G.S. Adams,E.I. Ivanov,J. Kuhn}. \\
Meson spectroscopy, mainly, the light meson spectroscopy gives the information of the nonperturbative regime of QCD. The basic difficulty to study the light meson spectrum is that mostly resonances do not come out as narrow, isolated peaks \cite{Brambilla-2}. Moreover, their large decay width make them difficult to identify experimentally  as well as to differentiate them as threshold or as interference effects, rather than pure resonances. Especially for scalar mesons, they are very difficult to identify due to large decay width, overlaps between resonances and background. Even, the scalar mesons have the same quantum numbers as of vacuum. Thus, the understanding the properties of scalar mesons may help us to understand the mechanism of symmetry breaking \cite{Brambilla-2,Yong-Liang Ma}. Various mesons in low-lying sector like $f_{0}(980)$, $a_{0}(980)$, $f_{0}(1500)$ etc. observed experimentally, but their specific structure and properties are not understood theoretically as well as experimentally. For example, the structure of the scalar states $f_{0}(1500)$ and $f_{0}(1710)$ have debated since long time \cite{Stanislaus Janowski,W.S. Carvalho,D. Parganlija,Frederic,L. S. Geng,Seungho Choe,Eberhard Klempt,A. Martínez Torres,R. Molina,C. Garcı́a-Recio,M. Ablikim,Geng,Branz}. The authors in \cite{Stanislaus Janowski,W.S. Carvalho,D. Parganlija} indicated them as mixed or rather pure glue state respectively. Whereas, the Ref. \cite{L. S. Geng,Seungho Choe,Eberhard Klempt,A. Martínez Torres,R. Molina,C. Garcı́a-Recio,M. Ablikim} indicated $f_{0}(1710)$ as a Vector-Vector molecular candidate and $f_{0}(1500)$ as a glueball. The status of these scalar states are still a open question. The study of these exotic states may give the answer to the basic questions of hadron masses, quark confinement, relevant degrees of freedom and interaction between constituents of the same \cite{Brambilla-2,Yong-Liang Ma}.\\

\begin{table*}[t]
\begin{center}
\caption{Experimentally observed states (in MeV) \cite{21}}
\label{tab-3}
\begin{tabular}{ cc c c c c}
\hline 
States &Mass& Widths &Possible & Compared \\
 & (MeV)&(MeV)& $I^G (J^{PC})$ &  dimesonic  \\
  & & &  & states  \\
  \hline 
$f_0(980)$&990$\pm$20 &40-100 &$0^+(0^{++})$ &$K-\bar{K}$   \\
$b_1(1235)$&1229$\pm$03 &$142\pm09 $ &$1^+(1^{+-})$ &$\eta-\bar{\rho}$   \\
$h_1(1380)$&1386$\pm$19 &91$\pm$30 &$0^-(1^{+-})$ &$K-\bar{K^*}$ \\
$a_0(1450)$&1474$\pm$19 &265$\pm$13 &$1^-(0^{++})$ &$\rho-\bar{\omega}$ \\
$f_0(1500)$&1505$\pm$06 &109$\pm$07 &$0^+(0^{++})$ &$\rho-\bar{\rho}$ \\
$f'_2(1525)$&1525$\pm$05 &76$\pm$10 &$0^+(2^{++})$ &$\rho-\bar{\rho}$ \\
$f_2(1565)$&1562$\pm$13 &134$\pm$08 &$0^+(2^{++})$ &$\omega-\bar{\omega}$ \\
$h_1(1595)$&1594$\pm$15 &384$\pm$60 &$0^-(1^{+-})$ &$\eta'-\bar{\omega}$ \\
$a_2(1700)$& $1722\pm16$ &194$\pm$40 &$1^-(2^{++})$ &$K^*-\bar{K^*}$ \\
$f_0(1710)$& $1722^{+06}_{-05}$ &135$\pm$07 &$0^+(0^{++})$ &$K^*-\bar{K^*}$ \\
$f_2(1810)$& 1815$\pm$12  &197$\pm$22 &$0^+(2^{++})$ &$\phi-\bar{\omega}$ \\

\hline 
\end{tabular}
\end{center}
\end{table*}

In the present study, we have investigated the molecular structures in the light flavour meson regime. Our dimesonic system consist of meson and antimeson ($q\overline{q}-\overline{q}q$). The masses of these states would be less than the sum of the masses of two constituent mesons \cite{17,22}. These loosely bound dimesonic system are similar to the deuteron like (Proton-neutron) bound state system. This approximation had been taken previously in Ref.\cite{22} and introduced as 'deusons'. Furthermore, the molecular like picture had been studied by Silvestre et. al. \cite{15} as diquonia.\\
The multiquark states were predicted and studied previously in MIT Bag model and in nonrelativistic potential models \cite{5,R.L. Jaffe,6}. The molecular like structure have been studied and proposed in various models like potential models \cite{17,22,15,11,14,18,Rui Zhang,Iwasaki2008,Bhattacharya2011}, chiral SU(3) quark model approach \cite{7}, gauge invariant model \cite{Tanja Branz}, Bethe-Salpeter equation approach \cite{26}, QCD sum rules  \cite{Jian-Rong Zhang}, an effective field theory approach \cite{C. Hidalgo-Duque} and nonperturbative chiral approach \cite{J. A. Oller}. We are using the potential model to predict the masses, digamma decay and decay width of dimesonic systems in variational scheme. The potential model study is successfully explain the heavy flavour sector of mesons as well as multiquark description of meson-antimeson or tetraquark states. The potential model are used to calculate the masses \cite{5,R.L. Jaffe,6,22,15,11,14} and decay properties of various multiquark states. Recently, the potential model has employed successfully to study various spectroscopic study of mesons \cite{Jing-Bin Liu,Jun-Kang He,Kh. Ablakulov,Vijaya kumar,Smruti Patel,Lesolle}. Whereas, various authors have successfully used non-relativistic potential model for the study of the molecular or tetra-quark structure \cite{14,Rui Zhang,D. Janc,J. Vijande,Marco Cardoso}.\\
In our model for study of dimesonic states, we are using the Hellmann potential with One Pion Exchange Potential (OPEP). The Hellmann potential is the superposition of coulomb and Yukawa potential. The superposed potential of coulomb and Yukawa was studied by Hellmann long time ago \cite{H-1,H-2,H-3}. There are several authors who applied this potential for the calculation of the bound state systems for different values of their strength parameters \cite{Roy,Janusz,Hamzavi,Nasser,Herbert}. All the phenomenological potential model approaches used by the authors, to study the mesons or multiquark spectra have dealt with complex interactions like,  coulombic, confinement, instanton induced interaction, color, flavour and spin dependency to accomplished the bound state properties. At very short distances, the bound state system is produce due to delicate attractive and repulsive interaction. The Hellmann potential is simply accomplished the complicated theoretical calculation for the cancellation of the attractive and repulsive interaction at short distances, causes the small binding energy of the bound state. The Hellmann potential with the OPEP used in the present model take care of short and long range behaviour of the interactions. As we aware that mesons are color neutral so we have not introduced the confinement potential for the dimesonic systems. It is well studied that the OPEP is mainly responsible for long range part of the strong nuclear force.  The concept of meson exchange interaction has been used since long time for the N-N interaction \cite{R. Machleidt,M. Naghdi} as well as for hadronic molecules \cite{22,19,20,23}.\\      
To test the potential and mechanism employed, we first apply it to study the state $f_{0}(980)$ which was found in literature to be interpreted as molecular state \cite{6,Rui Zhang}. Moreover, the molecular like picture has been supported by lattice QCD \cite{M. Alford}. We calculate the spectra of  $f_{0}(980)$ and obtained it's mass, binding energy and root mean square radius. The results shows fairly good agreement with experimental results. Thus we are convinced and apply the same methodology to other dimesonic molecular states. \\
The study of light meson sector needs relativistic treatment. We are incorporating relativistic correction to the kinetic energy term as well as to the meson-antimeson potential. The hydrogen like trial wave function is being used for this study. We have calculated the digamma decay width and decay width of the dimesonic systems.  The study of the digamma decay of scalars helps to distinguish among different scenarios for scalar meson structure \cite{18}. The digamma decay of the dimesonic systems are calculated using the wave function at the origin \cite{18}. For decay calculation, we adopt same formulation as used by Ref.\cite{Rui Zhang}, which was used to study the molecular picture of $1^{-+}$ exotic states.\\

The article is organized as follows. After the brief introduction, we present the theoretical framework for the study of dimesonic molecular states in semirelativistic approach with the Hellmann potential and OPE potential in section-II. The calculation of digamma decay and decay width are discussed in section-III. Then we presents the calculated results in section-IV and the  last section is dedicated for conclusion.

\section{Theoretical Framework}

\label{sec:sect2}
In the variational approach, we have solved the Schroed-
inger equation by using trial wave function. The variational parameter is being obtained by using the virial theorem. The (dimesonic) molecular system assumed to be consist meson-antimeson bound state. The Hamiltonian of the dimesonic system is given by \cite{9,8}
\begin{equation}
H=\sqrt{P^2+m_{a}^{2}}+\sqrt{P^2+m_{b}^2}+V(r)
\end{equation}
where $m_{a}$ and $m_{b}$ are masses of mesons, P is the relative
momentum of two mesons and V(r) is the molecular interaction potential
of the dimesonic system. We expand the kinetic energy term of the Hamiltonian up to $\cal{O}$($P^{6}$)

\begin{eqnarray}
K.E. &=& \frac{P^{2}}{2}\left(\frac{1}{m_{a}} +{\frac{1}{m_{b}}}\right)
- \frac{P^{4}}{8}\left(\frac{1}{m_{a}^{3}} + { \frac{1}{m_{b}^{3}}}\right)\nonumber \\ & & 
+ \frac{P^{6}}{16}\left(\frac{1}{m_{a}^{5}} +{ \frac{1}{m_{b}^{5}}}\right)+{\cal{O}}{(P^{8})}
\end{eqnarray}

V(r) is the molecular (meson-antimeson) potential, 
\begin{equation}
V\left(r\right)=V^{\left(0\right)}\left(r\right)+\left(\frac{1}{m_{a}}+\frac{1}{m_{b}}\right)V^{\left(1\right)}\left(r\right)+{\cal O}\left(\frac{1}{m^{2}}\right)
\end{equation}
where,

\begin{equation}
V^{(0)}(r)=V{(r_{12})}+V{}_{\pi}
\end{equation}
We consider the meson-antimeson interacting through the Hellmann potential. It is a superposition of the Coulomb and Yukawa potentials of the form \cite{H-1,H-2,H-3,Roy,Janusz,Hamzavi,Nasser}
\begin{equation}
V{(r_{12})}=-\frac{\alpha_{s}}{r_{12}}+B\frac{e^{-cr_{12}}}{r_{12}}
\end{equation}
Where $\alpha_{s}$ and B are the strength of potential, C is the screening parameter and $r_{12}$ is the relative coordinates between two meson and antimeson, the strength of coulomb potential is used as running coupling constant. While B is the Yukawa potential strength and it takes both positive and negative values. Ref. \cite{Janusz} had carried out a detailed study for different values of B and screening parameter C for low lying energy eigenvalues. For our dimesonic study, we have taken B positive. \\

The value of the $\alpha_{s}$  running coupling constant is determined through the model, namely
\begin{equation}
\alpha_{s}(M^{2}) = \frac{4\pi}{(11-\frac{2}{3}n_{f})ln\frac{M^{2}+ {M_{B}}^{2}}{\Lambda_{Q}^{2}}}
\end{equation}
where M = 2$m_{a}$ $m_{b}$/($m_{a}$+$m_{b}$), $M_{B}$ = 1 GeV, $\Lambda_{Q}$ = 0.413 GeV and $n_{f}$ is number of flavour \cite{Ebert2009}.

The nonperturbative form of $V^{\left(1\right)}\left(r\right)$ is not yet known, but leading order perturbation theory yields
\begin{equation}
V^{\left(1\right)}\left(r\right)=-C_{F}C_{A}\alpha_{s}^{2}/4r^{2}_{12};
\end{equation}
where $C_{F}=4/3$ and $C_{A}=3$ are the Casimir charges of the fundamental and adjoint representation, respectively \cite{Koma2006}. The relativistic mass correction is found to be similar to the Coulombic term of the static potential when applied to the charmonium and to be one-forth of the Coulombic term for bottomonium \cite{Koma2006}. Recently we have used this correction successfully to the study of $B_c$ meson \cite{9}.  

The long range one pion exchange potential (OPEP) is used based on the assumption that molecular like structure of multiquark system is being deuterium like structure of nucleon \cite{17,19}. The OPEP for the NN-interaction takes the form \cite{VijayR} 

\begin{eqnarray}
V{}_{\pi (NN)}&=&\frac{g^{2}{}_{8}}{4\pi}\frac{m_{\pi}}{3}(\tau{}_{i}\cdot\tau{}_{j})\nonumber \\ & &\left[T_{\pi}(r)S_{12} +  \left(Y_{\pi}(r) - \frac{4\pi}{m_{\pi}^{3}}\delta(r)\right)\sigma{}_{i}\cdot\sigma{}_{j}\right]
\end{eqnarray}
where  $g{}_{8}$ is a pion-nucleon coupling constant. {$\bf\tau$} and {$\bf\sigma$} are isospin, spin factors respectively.
where$ T_{\pi}(r)$, $Y_{\pi}(r)$ and $S_{12}$ are
\begin{equation}
T_{\pi}(r)=\left(1+\frac{3}{m_{\pi}r} + \frac{3}{m_{\pi}^{2}r^{2}}\right)\frac{e^{-m_{\pi}r}}{m_{\pi}r} 
\end{equation} 
\begin{equation}
Y_{\pi}(r)=\frac{e^{-m_{\pi}r}}{m_{\pi}r}
\end{equation}
\begin{equation}
S_{12}=3\sigma{}_{i}\cdot\hat{r}\sigma{}_{j}\cdot\hat{r}/r^{2}-\sigma{}_{i}\cdot\sigma{}_{j}
\end{equation}
The exchange meson having its own internal structure, to accomplished the finite size effect \cite{VijayR}, the usual form factor is appear due to the dressing of quarks and is assumed to be  proportional to the exchange meson mass \cite{19}.
\begin{equation}
T_{\Lambda_{\pi}}(r)=\left(1+\frac{3}{{\Lambda_{\pi}}r} + \frac{3}{{\Lambda_{\pi}}^{2}r^{2}}\right)\frac{e^{-{\Lambda_{\pi}}r}}{{\Lambda_{\pi}}r} 
\end{equation} 
\begin{equation}
Y_{\Lambda_{\pi}}(r)=\frac{e^{-{\Lambda_{\pi}}r}}{{\Lambda_{\pi}}r}
\end{equation}
The functions $T(r)$ and $Y(r)$ with the finite size effects are takes the form
\begin{equation}
Y(r)=\frac{\Lambda_{\pi}^{2}}{\Lambda_{\pi}^{2}-m_{\pi}^{2}}\left[Y_{\pi}(r) - \frac{\Lambda_{\pi}^{3}}{m_{\pi}^{3}}Y_{\Lambda_{\pi}}(r)\right]
\end{equation}
\begin{equation}
T(r)=\frac{\Lambda_{\pi}^{2}}{\Lambda_{\pi}^{2}-m_{\pi}^{2}}\left[T_{\pi}(r) - \frac{\Lambda_{\pi}^{3}}{m_{\pi}^{3}}T_{\Lambda_{\pi}}(r)\right]
\end{equation} 
The more detail derivation of the potential can found in the Ref. \cite{VijayR}. The 
$S_{12}$ is usual tensor operator, plays prominent role in NN-interaction. The matrix element of the tensor operator for L=0  vanishes \cite{23}. If two hadrons are in an L=0 (ground state) state, the term with tensor operator vanishes as in our dimesonic case. Thus, the OPEP takes the form

\begin{eqnarray}
V{}_{\pi}&=&\frac{1}{3}\frac{g^{2}{}_{8}}{4\pi}\left(\frac{m{}_{\pi}^{2}}{4m_{a}m_{b}}\right)\left(\tau{}_{i}\cdot\tau{}_{j}\right) 
\left(\sigma{}_{i}\cdot\sigma{}_{j}\right)\nonumber \\ & & 
\left(\frac{e{}^{- m_{\pi}r_{ij}}}{r_{ij}}-\left(\frac{\Lambda{}_{\pi}}{m_{\pi}}\right)^{2}\frac{e{}^{-\Lambda_{\pi}r_{ij}}}{r_{ij}}\right)
\end{eqnarray}

where  $g{}_{8}$ = 0.69 is a meson pion coupling constant, $\Lambda{}_{\pi}=km{}_{\pi}$, whereas, $m_{a}$ and $m_{b}$ are constituents masses, $m_{\pi} = 0.134$ GeV and  $k = 2.2$. For PV states the values of spin-isospin factor have been taken as $\left(\tau{}_{i}\cdot\tau{}_{j}\right)\left(\sigma{}_{i}\cdot\sigma{}_{j}\right)\ $ = -3,1 for I = 0,1. While for VV states  values taken as $\left(\tau{}_{i}\cdot\tau{}_{j}\right)\left(\sigma{}_{i}\cdot\sigma{}_{j}\right)\ $ = -6,-3,3 for isospin I = 0 and spin S = 0,1,2 respectively whereas for isospin I = 1 and spin S = 0,1,2 it takes values $\left(\tau{}_{i}\cdot\tau{}_{j}\right)\left(\sigma{}_{i}\cdot\sigma{}_{j}\right)\ $ = 2,1,-1. The values of spin-isospin factor are from Ref.\cite{22} and the other parameters used in pion exchange potential are from Ref. \cite{19,20,23}.

For the hyperfine splitting, the spin dependent interaction potential added perturbatively and takes the form

\begin{equation}
V_{SD}=\frac{8}{9}\frac{\alpha_{s}}{m_{a}m_{b}} {\bf S_{1}\cdot S_{2}} \left|\psi(0)\right|^{2}
\end{equation}
 The spin factor ${\bf S_{1}\cdot S_{2}}$ can be found by general formula ${\bf S_{1}\cdot S_{2}}=\frac{1}{2}[({\bf S_{1}}+{\bf S_{2}})^{2}-{\bf S_{1}}^{2}-{\bf S_{2}}^{2}]$ \cite{11} \\

We have used the hydoginic like trial wave function such as,
\begin{equation}
R_{nl}(r)=\left(\frac{\mu{}^{3}(n-l-1)!}{2n(n+l)^{3}!}\right)^{\frac{1}{2}}\left(\mu r\right)^{l}e{}^{\frac{-\mu r}{2}}L_{n-l-1}^{2l+1}(\mu r)
\end{equation}

\begin{table*}[]
\begin{center}
\caption{Masses of mesons (in MeV) \cite{21}}
\label{tab-3}
\begin{tabular}{ c c c c c c c c c}
\hline 
Meson  & $K{}^{+}$ & $K{}^{0}$ & $\eta$ & $\eta'$ & $\rho$ & $\omega$ &$K{}^{*}$ &  $\phi$ \tabularnewline
Mass & 493.6 & 497.6 & 547.8 & 957.7 & 775.4 & 782.6  & 895.9& 1019.4  \tabularnewline
\hline 
\end{tabular}
\end{center}
\end{table*}

\noindent where $\mu$ is the variational parameter and 
$L_{n-l-1}^{2l+1}(\mu r_{12})$ is the Laguerre polynomial.
The ground state mass of the low-lying dimesonic states is calculated by obtaining the expectation value of the Hamiltonian and the variational parameter($\mu$) is obtained for each state by using the virial theorem \cite{9,8}.

\begin{equation}
H\psi=E\psi \\ and \\ \langle K.E.\rangle = \frac{1}{2}\left <\frac{  r  d V(r)}{dr}\right>
\end{equation}
We have included the kinetic energy term upto $P^6$ and potential with correction (Eq(3)) for the determination of $\mu$. We fix color screening parameter C = 0.35 GeV for all combination and B = 0.4 for PP and PV-states and B = 1.2 for VV-states while strength of coulomb potential $\alpha(M^{2})$ is calculated as per Eq(6). The experimental (PDG) masses of the mesons are used for the present study, tabulated in Table-(II) \cite{21}.

The angular momentum,parity, spin, isospin,  all are conserved in strong interaction, being good quantum number for  dimesonic system. The orbital angular momentum of meson and antimeson are $L_{1}$ and $L_{2}$ respectively. In same way spin denote by $S_{1}$, $S_{2}$ and isospin $I_{1}$, $I_{2}$. Employing the coupling rules, one have the relative orbital momentum and total spin of system $L_{12}$ and $S_{12}$ whereas the total angular momentum to be {\bf{J}} = { $\bf L_{12}$}+{  $\bf S_{12}$} whereas {\bf I} = {$\bf I_1$}+{ $\bf I_2$}. The  charge conjugation of the two particle system (meson-antimeson) is given by C = $(-1)^{L_{12}+S{12}}$ \cite{40} while the parity be P = $P_{1}$ $P_{2}$ $(-1)^{L_{12}}$ \cite{15} whereas G-parity is defined as G = $(-1)^{L_{12}+S_{12}+I}$. The quantum numbers for conventional mesons as well as for exotic states have been discussed in Ref.\cite{15,40}.\\ 
The one pion exchange potential (OPEP) for dimesonic system in Eq.(8) is spin-isospin dependent.  T$\ddot{o}$rnqvist in \cite{22} had discussed the spin-isospin dependency of OPEP for deusons.
The spin-isospin factor gives influence to the potential and decides whether channel becomes attractive or repulsive. For VV-states we got repulsive channel for (S,I) = (0,1)(1,1)(2,0) while  attractive channel for (S,I) = (0,0)(1,0)(2,1). Whereas in the case of PV-states the channel becomes attractive for (S,I) = (1,0) while repulsive for (S,I) = (1,1). Due to parity violation, the two Pseudoscalar could not be bound by a Pseudoscalar \cite{17,22}. Thus, we could not consider OPEP for PP-states. For such PP-states, the Hellmann potential given in Eq.(5) is being used with relativistic correction. The calculated masses are close to experimental measurements(PDG), tabulated in Table-(III). 

We have used Pseudoscalar and Vactor meson for the dimesonic system, having possible combinations (i) Pseudoscalar-Pseudoscalar(PP-state) (ii) Pseudoscalar-Vector (PV-state) (iii) Vector-Vector (VV-state). \\
\section{Di-gamma width and decay width}
The digamma decay of the dimesonic molecules are estimated, using the wave function at the origin, in analogy to the two photon decay of parapositronium \cite{18}. 
The di-gamma width for all mass combinations is given by 
\begin{equation}
\Gamma_{\gamma\gamma}=2 \xi^2 \frac{\pi \alpha^2}{m_{a}m_{b}} \left|\psi(0)\right|^2
\end{equation}
 $\alpha=e^{2}/4\pi$ indicate fine-structure constant, the
factor $\xi$ =$\frac{1}{\sqrt{2}}$, $m_{a}$ and $m_{b}$ are the masses of constituent mesons. \\

The decay width is calculated using the wave function at the origin. The dimesonic molecular  state (constituent mesons a and b) is decays into mesons c and d. In the decay through meson exchange, the amplitude is proportional to the wave function square at the origin \cite{Rui Zhang}. The  formula of $\left|\psi(0)\right|^2 $ given by \cite{Rui Zhang,Quigg,X.-Q. Li}

\begin{equation}
\ \left|\psi(0)\right|^2 = 2 M <\psi\mid dV/dr \mid \psi>
\end{equation}
with M as reduced mass of the system, $\psi(0)$ includes the effects of all ranges in r.

The decay width is 

\begin{equation}
\Gamma =  \frac{\ \left|\psi(0)\right|^{2}l}{16\pi m_{m}^{3}} \left|{\cal{M}}\right|^{2}
\end{equation}

 Where $m_{m}$ is the mass of dimesonic molecule and l is the magnitude of 3-momentum of the decay product and is given by 
 
\begin{eqnarray}
l^{2} & = & \frac{({m_{m}^{4}+m_{c}^{4}+{m_{d}^{4}}})}{4m_{m}^{4}} \nonumber \\ & &  - \frac{(m_{m}^{2}m_{c}^{2}+m_{m}^{2}m_{d}^{2}+m_{c}^{2}m_{d}^{2})}{2m_{m}^{4}}
\end{eqnarray}

$m_{a}$, $m_{b}$ are masses of constituent mesons and $m_{c}$, $m_{d}$ are  masses of product mesons. 
whereas $\cal{M}$ is the amplitude 

\begin{equation}
{\cal{M}}=\frac{ \alpha_{s}^{2}}{q^{2}-m_{q}^{2}}\left(\frac{\Lambda^{2}-m_{q}^{2}}{\Lambda^{2}-{q}^{2}}\right)
\end{equation}

$m_{q}$ and q are the mass and 3-momentum of exchange mesons (for present work, we consider only a pion). $\Lambda$ is the free parameter which care of the off-shell effects at the vertices because of the internal structure of the mesons. 
The digamma decay widths are tabulated in Table-III along with masses. The decay width of compared states as well as our predicated states are tabulated in Table-IV.

\begin{table*}
\begin{center}
\caption {Mass spectra, binding energy, root mean square radius and  digamma decay width of dimesonic ($q\overline{q}-\overline{q}q$) systems  with their $J^{PC}$ values.}
\label{tab-1}
\begin{tabular}{ c c c c c c c c c c c}
\hline 
System & $I{}^{G}(J^{PC})$ & $\mu$ & R(0) & $\sqrt{<r^{2}>}$ & B.E. & Mass & Expt. \cite{21} & $\Gamma_{\gamma\gamma}$ & Exp. \cite{21} & State\tabularnewline
 
 &  & (GeV) & $(GeV{}^{\frac{3}{2}})$  & (fm) & (MeV) & (GeV) & (GeV) & KeV & KeV & \tabularnewline
\hline
PP-states &&&&&&&&&&\\

$K-\overline{K}$  & $0{}^{+}$ $(0{}^{++})$ & 0.219 & 0.072 & 11.63 & -20.87  & 0.974 & $0.990$ & 0.285 & $0.29_{-0.06}^{+0.07}$ & $f{}_{0}(980)$\tabularnewline

&&&&&&&$\pm0.020$&&& \tabularnewline

$\eta-\overline{\eta}$ & $0{}^{+}$ $(0{}^{++})$ & 0.265 & 0.096 & 09.63 & -26.24 & 1.069 & & 0.414 & & \tabularnewline

$\eta-\overline{\eta}'$ & $0{}^{+}$ $(0{}^{++})$ & 0.409 & 0.185 & 06.24 & -46.75 & 1.458 &  & 1.526 & &  \tabularnewline
 
$\eta'-\overline{\eta}'$& $0{}^{+}$ $(0{}^{++})$  & 0.628 & 0.352 & 04.06 & -85.22  & 1.830 &  & 1.803 &  &  \tabularnewline
\hline
PV-states &&&&&&&&&&\\
$\eta-\overline{\rho}$ & $1{}^{+}$ $(1{}^{+-})$ & 0.210 & 0.068 & 12.12 & -63.82  & 1.259 & $1.229$ & 0.208 & & $b_{1}(1235)$  \tabularnewline

&&&&&&&$\pm0.003$&&& \tabularnewline

$\eta-\overline{\omega}$ & $0{}^{-}$ $(1{}^{+-})$ & 0.209 & 0.067 & 12.20 & -63.22 & 1.267 & & 0.203 & &  \tabularnewline

$K-\overline{K}^{*}$ &$0{}^{-}$ $(1{}^{+-})$ & 0.206 & 0.066 & 12.36 & -62.60  & 1.330 & $1.386$ & 0.237 & & $h_{1}(1380)$  \tabularnewline

&&&&&&&$\pm0.019$&&& \tabularnewline

$K-\overline{K}^{*}$ & $1{}^{+}$ $(1{}^{+-})$  & 0.209 & 0.067 & 12.19 & -63.51 & 1.330 &  & 0.247 & &  \tabularnewline

$\eta-\overline{\phi}$ & $0{}^{-}$ $(1{}^{+-})$ & 0.253 & 0.090 & 10.08 & -73.50  & 1.493 &  & 0.361 & &  \tabularnewline

$\eta'-\overline{\rho}$ & $1{}^{+}$ $(1{}^{+-})$ & 0.351 & 0.147 & 07.26 & -94.25  & 1.639 &  & 0.316 & &  \tabularnewline
 
$\eta'-\overline{\omega}$& $0{}^{-}$ $(1{}^{+-})$ & 0.350 & 0.146 & 07.30 & -93.45 & 1.646 & $1.594$ & 0.312 & &  $h_{1}(1595)$ \tabularnewline

&&&&&&&$\pm0.015$&&& \tabularnewline
 
$\eta'-\overline{\phi}$& $0{}^{-}$ $(1{}^{+-})$ & 0.429 & 0.198 & 05.95 & -107.4  & 1.869 & & 0.574	 & &  \tabularnewline

\hline 
VV-states &&&&&&&&&&\\
$\rho-\overline{\rho}$ & $0{}^{+}$ $(0{}^{++})$ & 0.192 & 0.059 & 13.27 & -55.39 & 1.489 & $1.505$ & 0.079 & & $f{}_{0}(1500)$\tabularnewline

&&&&&&&$\pm0.006$&&& \tabularnewline

$\rho-\overline{\rho}$ & $0{}^{-}$ $(1{}^{+-})$ & 0.197 & 0.062 & 12.92 & -55.00 & 1.492 &  & 0.085 & & \tabularnewline
 
$\rho-\overline{\rho}$ & $0{}^{+}$ $(2{}^{++})$ & 0.194 & 0.060 & 13.15 &-54.10 & 1.500 & $1.525$ & 0.081 & $0.081$ & $f{}_{2}'(1525)$ \tabularnewline

&&&&&&&$\pm0.005$&&$\pm0.009$& \tabularnewline
 
$\rho-\overline{\rho}$ & $1{}^{-}$ $(0{}^{++})$ & 0.194 & 0.060 & 13.11 & -54.25 & 1.490 &  & 0.082 & & \tabularnewline
 
$\rho-\overline{\rho}$ & $1{}^{+}$ $(1{}^{+-})$ & 0.195 & 0.061 & 13.07 & -54.40 & 1.493 &  & 0.082 & & \tabularnewline

$\rho-\overline{\rho}$ & $1^{-}$ $(2{}^{++})$ & 0.196 & 0.061 & 13.00 & -54.70 & 1.499 &  & 0.084 & &  \tabularnewline
\hline 
$\omega-\overline{\omega}$ & $0{}^{+}$ $(0{}^{++})$ & 0.201 & 0.064 & 12.67 & -55.77 & 1.502 &  & 0.089 & & \tabularnewline

$\omega-\overline{\omega}$ & $0{}^{-}$ $(1{}^{+-})$ & 0.199 & 0.063 & 12.79 & -55.31 & 1.506 &  & 0.086 & & \tabularnewline
 
$\omega-\overline{\omega}$ & $0{}^{+}$ $(2{}^{++})$ & 0.196 & 0.061 & 13.01 & -54.40 & 1.514 & $1.562$  & 0.082 & $0.70 $ & $f{}_{2}(1565)$  \tabularnewline

&&&&&&&$\pm0.013$&&$\pm0.14$& \tabularnewline
\hline 
$K^{*}-\overline{K}^{*}$ & $0{}^{+}$ $(0{}^{++})$ & 0.231 & 0.078 & 11.05 & -59.34 & 1.724 &  $1.722_{-0.005}^{+0.006}$ & 0.102 &  & $f{}_{0}(1710)$ \tabularnewline

$K^{*}-\overline{K}^{*}$ & $0{}^{-}$ $(1{}^{+-})$ & 0.229 & 0.077 & 11.14 & -58.89 & 1.729 &  & 0.100 & &  \tabularnewline

$K^{*}-\overline{K}^{*}$ & $0{}^{+}$ $(2{}^{++})$ & 0.225 & 0.075 & 11.32 & -58.01 & 1.730 &  & 0.095 & &  \tabularnewline
 
$K^{*}-\overline{K}^{*}$ & $1{}^{-}$ $(0{}^{++})$ & 0.226 & 0.076 & 11.29 & -58.16 & 1.726 &  & 0.096 & &  \tabularnewline
 
$K^{*}-\overline{K}^{*}$ & $1{}^{+}$ $(1{}^{+-})$ & 0.226 & 0.076 & 11.26 & -58.31 & 1.730 &  & 0.096 & &  \tabularnewline

$K^{*}-\overline{K}^{*}$ & $1^{-}$ $(2{}^{++})$ & 0.228 & 0.077 & 11.20 & -58.60 & 1.736 & $1.722 $ & 0.098 & $0.30$ &  $a{}_{2}(1700)$  \tabularnewline

&&&&&&&$\pm0.016$&&$\pm0.05$& \tabularnewline
\hline 
$\phi-\overline{\phi}$ & $0{}^{+}$ $(0{}^{++})$ & 0.255 & 0.091 & 10.00 & -60.91 & 1.970 &  & 0.106 & &  \tabularnewline

$\phi-\overline{\phi}$ & $0{}^{-}$ $(1{}^{+-})$ & 0.253 & 0.090 & 10.07 & -60.49 & 1.974 &  & 0.104 & &  \tabularnewline
 
$\phi-\overline{\phi}$ & $0{}^{+}$ $(2{}^{++})$ & 0.250 & 0.088 & 10.22 & -59.68 & 1.982 &  & 0.100 & &  \tabularnewline
\hline 
$\rho-\overline{\omega}$ & $1{}^{-}$ $(0{}^{++})$ & 0.196 & 0.061 & 13.04 & -54.40 & 1.497 & $ 1.474$ & 0.081 & &  $a{}_{0}(1450)$ \tabularnewline

&&&&&&&$\pm0.019$&&& \tabularnewline

$\rho-\overline{\omega}$ & $1{}^{+}$ $(1{}^{+-})$ & 0.196 & 0.061 & 13.00 & -54.55 & 1.500 &  & 0.082 & &  \tabularnewline
 
$\rho-\overline{\omega}$ & $1^{-}$ $(2{}^{++})$ & 0.197 & 0.062 & 12.93 & -54.85 & 1.506 & & 0.084 & &  \tabularnewline
 \hline
$\phi-\overline{\rho}$ & $1{}^{-}$ $(0{}^{++})$ & 0.222 & 0.074 & 11.47 & -57.82 & 1.730 &  & 0.070 & &  \tabularnewline
 
$\phi-\overline{\rho}$ & $1{}^{+}$ $(1{}^{+-})$ & 0.223 & 0.074 & 11.44 & -57.96 & 1.733 &  & 0.071 & & \tabularnewline
 
$\phi-\overline{\rho}$ & $1^{-}$ $(2{}^{++})$ & 0.224 & 0.075 & 11.38 & -58.25 & 1.740 &  & 0.072 & & \tabularnewline
 \hline
$\phi-\overline{\omega}$ & $0{}^{+}$ $(0{}^{++})$ & 0.228 & 0.077 & 11.17 & -59.10 & 1.735 & & 0.076 & &  \tabularnewline
 
$\phi-\overline{\omega}$ & $0{}^{-}$ $(1{}^{+-})$ & 0.226 & 0.076 & 11.26 & -58.65 & 1.739 &  & 0.074 & & \tabularnewline
 
$\phi-\overline{\omega}$ & $0{}^{+}$ $(2{}^{++})$ & 0.223 & 0.074 & 11.44 & -57.79 & 1.747 & $1.815 $ & 0.071 & & $ f_{2}(1810)$ \tabularnewline

&&&&&&&$\pm0.012$&&& \tabularnewline
\hline 

\end{tabular}
\end{center}
\end{table*}

\section{Results and Discussion}
By using the Hellmann potential, OPE potential and relativistic correction, we have solved the Schroedinger equation to extract the masses, binding energy, decay widths and  digamma decay widths for dimesonic states. In various combination of dimesonic states like PP, PV and VV- states, we have calculated the possible molecular like structure in light flavoured sector. Our calculated results are tabulated in Table (III - IV).\\
To describe molecular picture of dimesonic states, we have incorporated the short range and long range behaviour of the potential. As the constituent of the molecule (meson) itself color neutral and with the small binding energy and large radius (r in fm) in comparison to constituent size, we do not need to introduced the confinement potential and so  we have not dealt with fundamental quark gluon condense state in the present study. Thus, we have not incorporated any mixing scheme of gluonia and quarkonia. Still, the results obtained in present study have reasonably good agreement with experimental measurements and explain the molecular picture well.\\
We have expanded the kinetic energy term up to $\cal{O}$($P^{6}$). In the series expansion of the kinetic energy, For $v<<c$, the effect of the higher order term of the momentum $P^{2n}$ ($n>2$) is negligible, even more, the higher order term has poor convergence. While the expansion term up to $P^{4}$ does not have a lower bound. So, the usable expansion to incorporate the relativistic effect is being up to $P^{6}$ \cite{VijayR}. We have dealt with the systems which have constituent masses below  2 GeV. For the lighter constituents, the effect of the momentum has small. As the $v<c$ tends to $v<<c$, the effect of the higher order terms contributes less than $1 {\%}$. The effects of the higher order terms (up to $P^{6}$) are very small, still, it  has variation with increasing masses and contribute ($v\rightarrow c$) to the net kinetic energy of the system. Therefore, it justified to incorporate the expansion up to $P^{6}$.\\
The  correction used in the potential, have its dominant effect on potential energy. In our calculation, the effect of the correction to the potential energy part is increasing as mass of the system decreasing, and, it is about to be $25 - 40 \%$. The effect of correction to potential has approximate  $38\%$ in the lightest dimesonic system ($K-\overline{K}$ ) of this work, while, the heaviest system $\phi-\overline{\phi}$ has about $28\%$ contribution.
We have calculated the partial decay width as per Eq.(22). The parameter $\Lambda$ is assumed to be  proportional to the mass of the dimesonic state, $\Lambda$ = $\cal{K}$ $m_{m}$. We have chosen the Yukawa strength parameter and color screening parameter B = 0.4 and C = 0.1 respectively to find  mass spectra for $K-\overline{K}$ molecule which believed to be $f_{0}(980)$. For the decay calculation of $K-\overline{K}$, we took $\cal{K}$ = 1.262. Our calculated results for mass, digamma and decay width are close to the PDG values  \cite{21}. Thus, we fix these parameters for all PP and PV-states calculation while for VV-state, we took B = 1.2. For the dimesonic states having masses between 1 GeV to 1.6 GeV, the constant $\cal{K}$ = 1.085, whereas, for the states between 1.6 GeV to 2 GeV, $\cal{K}$ = 0.931.      
In PP-states, we used the Hellmann potential with relativistic correction. Due to parity violation,  the OPEP can not be apply to the PP-states calculation.\\
   
${m_{m}}_{f_{0}(980)}$ = 0.974 GeV ; $\Gamma_{\gamma\gamma}$ = 0.28 KeV ; 

$\Gamma_{(f_{0}(980))}$ = 94 MeV \\

Our results for $f_{0}(980)_{k\overline{K}}$ is in good agreement with suggested mass and decay width ($ 0.982\pm0.003 $ GeV and $80\pm10$ MeV) of Ref.\cite{D. Barberis}. The calculated digamma width of $f_{0}(980)$ is in close agreement with other theoretical predictions \cite{Tanja Branz,D. Barberis,30,31,32,33}. The $\pi\pi$ decay mode of $f_{0}(980)$ is dominant and also agree with Ref \cite{21}. For the molecular state, one should have small binding energy and large radius (compared to constituent energy and radius). Our results satisfied these conditions. The mass and decay properties are close to the experimental measurements. Thus, we are employing same methodology to calculate the properties of all the meson-antimeson combinations. The computed results are tabulated in Table-3 $\&$ 4.\\

Pseudoscalar-Vector (PV) states :- \\ 
In the PV-states, the  $ \pi V $ system  being the lightest  dimesonic state to be possible. In the present work, we have not considered $ \pi V $ systems. The pion is too light as a constituent for dimesonic states which interact through exchange of pion itself. The pion as a constituent, carrying the large kinetic energy and it is difficult to overcome by potential energy to bound a molecule \cite{22}. With the reasonably close mass spectra and quantum numbers, we have found three states $b_{1}(1235)$, $h_{1}(1380)$ and $h_{1}(1595)$ which could compared with $\eta-\overline{\rho}$, $K-\overline{K}^{*}$, $\eta'-\overline{\omega}$ dimesonic states respectively. The mass of  $b_{1}(1235)_{(\eta\overline{\rho})}$ has fairly near to experimental observation.\\
 
${m_{m}}_{b_{1}(1235)}$ = 1.259 GeV ; $\Gamma_{\gamma\gamma}$ = 0.20 KeV ;

$\Gamma_{(b_{1}(1235))}$ = 132 MeV\\

the $\pi\omega$ is experimentally observed dominant decay mode and our results are in good agreement.

For $h_{1}(1380)$

${m_{m}}_{h_{1}(1380)}$ = 1.330 GeV ; $\Gamma_{\gamma\gamma}$ = 0.23 KeV ;

$\Gamma_{(h_{1}(1380))}$ = 268 MeV

For $h_{1}(1595)$

${m_{m}}_{h_{1}(1595)}$ = 1.646 GeV ; $\Gamma_{\gamma\gamma}$ = 0.31 KeV ;

$\Gamma_{(h_{1}(1595))}$ = 63 MeV\\

In the case of the  $h_{1}(1380)_{(K\overline{K*})}$, our calculated decay widths are overestimated  . Moreover, the mass of the state is underestimated around 50 MeV. For the  $f_{1}(1595)_{(\eta'\overline{\omega})}$, the  decay width is unacceptably underestimated with experimental measurement with $\omega\eta$ observed decay mode. The mass of the state is also away from PDG values around 40 MeV. Thus, the obtained results indicates the rejection of molecular interpretation of both the states. Thus, in the PV-states calculations, the $b_{1}(1235)$ is the only the state, has been found the strong candidature for dimesonic molecule. \\

The Vector-Vector (VV) states :- \\
In the VV combination, the dependency of one pion interaction potential on spin-isospin factor gives  possibilities of the numbers of combinations for dimesonic  molecular like states \cite{22,23}. We have already discussed in the previous section about the matrix element, attractive and repulsive channels of the spin-isospin factors.  We have found bound states in repulsive spin-isospin channels, shows the dominance of hellmann potential and the mass correction over pion exchange. This can be clearly seen from binding energy tabulated in Table-IV for a state of same spin with different isospin. With different spin-isospin  combination,  we have compared experimentally observed states with our dimesonic molecular states, with comparable mass spectra and quantum numbers. The states  $f_{0}(1500)$, $f{}_{2}'(1525)$, $f{}_{2}(1565)$, $a_{0}(1450)$, $f_{0}(1710)$, $a_{2}(1700)$, $f_{2}(18 \\ 10)$ compared with different vector-vector combination, shown in Table-III.
 
The $f_{0}(1500)$ is being compared as  $\rho-\overline{\rho}$  molecule with (S,I)=(0,0), falling in 1500 GeV mass regime. The computed mass and decay widths are comparable with experimental values \cite{24} and with Ref. \cite{24}, suggested mass $1.522\pm0.005$ GeV and decay width $108\pm8 $ MeV. In Ref. \cite{D. Barberis} suggested decay width $ 131\pm15$ MeV, also in Ref. \cite{43,43-1,43-2}. We have found the $\pi\pi$ decay mode is dominant. Experimentally two photon decay has not been seen yet. 

${m_{m}}_{f_{0}(1500)}$ = 1.489 GeV ; $\Gamma_{\gamma\gamma}$ = 0.079 KeV ;

$\Gamma_{(h_{1}(1500))}$ = 110 MeV \\

The state  $f{}_{2}'(1525)$  has been compared with $\rho-\overline{\rho}$ dimesonic state, with (S,I)=(2,0). Our calculated decay width for $f{}_{2}'(1525)$ is consistent with Ref.\cite{21,35} whereas digamma width has agreement with Ref. \cite{21,32,34}.

${m_{m}}_{f_{2}'(1525)}$ = 1.500 GeV ; $\Gamma_{\gamma\gamma}$ = 0.081 KeV ;

$\Gamma_{(f_{2}'(1525))}$ = 82 MeV \\

The state $f{}_{2}(1565)$ is compared with $\omega-\overline{\omega}$ molecule. The calculated mass of this state is underestimated around 40 MeV with the PDG value \cite{21}, and, the calculated digamma decay width  is being underestimated about one ninth of the recent PDG value \cite{21}. More even, the calculated decay width for observed decay modes are also far away from PDG values \cite{21}. \\ 

for $f{}_{2}(1565)$

${m_{m}}_{f_{2}(1565)}$ = 1.514 GeV ; $\Gamma_{\gamma\gamma}$ = 0.082 KeV ;

$\Gamma_{(f_{2}(1565))}$ = 81 MeV \\

The states $f{}_{0}(1710)$ and $a{}_{2}(1700)$ are compared with $K^{*}-\overline{K}^{*}$ dimesonic molecule with spin state S=0,2 respectively. The mass of $f{}_{0}(1710)_{k^{*}\overline{K^{*}}}$ has  good agreement with PDG \cite{21} and relatively near to Ref. \cite{D. Barberis}  ($1.750\pm0.020$ GeV) and Ref.\cite{25,37}. Even, the computed decay width consistent with current experiment  \cite{21} and with Ref. \cite{25,44}. \\

${m_{m}}_{f_{0}(1710)}$ = 1.724 GeV ; $\Gamma_{\gamma\gamma}$ = 0.10 KeV ;

$\Gamma_{(f_{0}(1710))}$ = 143 MeV \\

The mass of $a{}_{2}(1700)_{k^{*}\overline{K^{*}}}$ is in excellent agreement with experimental PDG value \cite{21}.  The digamma decay is found one third of the PDG value ($0.30\pm0.05 KeV $) \cite{21}, but, the decay width is in good agreement with experimental values for observed decay modes \cite{21}.\\

${m_{m}}_{a_{2}(1700)}$ = 1.736 GeV ; $\Gamma_{\gamma\gamma}$ = 0.098 KeV ;

$\Gamma_{(a_{2}(1700))}$ = 181 MeV \\

Further more, we compared $\rho-\overline{\omega}$ and $\phi-\overline{\omega}$ dimesonic molecules with the states $a{}_{0}(1450)$ and $f{}_{2}(1810)$ respectively. The mass of $a_{0}(1450)_{\rho\overline{\omega}}$ is in agreement with PDG value \cite{21}. Whereas the decay width of $a_{0}(1450)_{\rho\overline{\omega}}$ has been found far away with PDG value \cite{21}, enforce us to rule out it as a dimesonic molecule. \\

${m_{m}}_{a_{0}(1450)}$ = 1.497 GeV ; $\Gamma_{\gamma\gamma}$ = 0.081 KeV ; 

$\Gamma_{(a_{0}(1450))}$ = 94 MeV \\

Whereas, The mass of the $f{}_{2}(1810)_{\phi\overline{\omega}}$  is off around 50 MeV  with experimental measurement \cite{21}, but the decay width are fairly comparable to PDG values in observed decay modes.  \\

${m_{m}}_{f_{2}(1810)}$ = 1.747 GeV ; $\Gamma_{\gamma\gamma}$ = 0.071 KeV ; 

$\Gamma_{(f_{2}(1810))}$ = 196 MeV \\

A.V. Anisovich et.al. \cite{46} made a remark on the status of the state $f{}_{2}(1810)$  such that the state may be confuse with $f{}_{0}(1790)$. \cite{46} has made  note, if there was not any confusion between $f{}_{2}(1810)$ and $f{}_{0}(1790)$ there should missing $0^{++}$ state.
As we observed throughout our calculated results, all states (except few states) are underestimated approximate 20 to 40 MeV. If we consider it as limitation of our model, we can compare the state $f_{0}(1790)$ with $\phi-\overline{\omega}$, with (S,I)=(0,0). So it may be possible that these two states $f{}_{2}(1810)$ and $f{}_{0}(1790)$ having molecular structure of $\phi-\overline{\omega}$ with spin state S = 0,2 respectively. But, the experimental status of both the states are not confirmed. Thus, it needs more attention from theoretically as well experimentally. Whereas, our results favoured the molecular picture.  \\
The status of the scalar states $f_{0}(1500)$ and $f_{0}(1710)$  as a molecule, guleball or glue mixed state have debated since long time \cite{Stanislaus Janowski,W.S. Carvalho,D. Parganlija,Frederic,L. S. Geng,Seungho Choe,Eberhard Klempt,A. Martínez Torres,R. Molina,C. Garcı́a-Recio,M. Ablikim,Geng,Branz}. The glueball and quarkonia with same quantum number and nearly close mass, may have interfered with each other and formed new state. If the one bare glueball and one or more quarkonia interfere, needs the mixing scheme to explain them. This interference or mixing of the state could be solved in mass matrix formalism  or in a linear combination of pure quarkonia and gluonia states by finding the linear coefficients which mainly depends on decay properties \cite{W.S. Carvalho}.
The authors of Ref. \cite{Stanislaus Janowski} have predicated  $f_{0}(1500)$ as a glue mixture with $s\overline{s}$ component and $f_{0}(1710)$ as  predominant a pure glueball, also in Ref \cite{W.S. Carvalho,D. Parganlija}. In Ref. \cite{Frederic} has indicated the $f_{0}(1710)$ as glueball candidate. Whereas the Ref. \cite{L. S. Geng,Seungho Choe,Eberhard Klempt,A. Martínez Torres,R. Molina,C. Garcı́a-Recio,M. Ablikim} have indicated $f_{0}(1710)$ as a Vector-Vector molecular candidate and $f_{0}(1500)$ as a glueball. So far, as per the present literatures, it has remained unclear and debatable the structure of the $f_{0}(1500)$ and  $f_{0}(1710)$. Moreover, in the recent review of particle data \cite{21} have noted on the status of the scalar states, in which $f_{0} (1710)$ together with $f_{2}′(1525)$ were interpreted as bound systems of two vector mesons \cite{21,R. Molina}. The molecular picture could be tested in radiative $J/\psi$ decays as well as radiative decays of the states themselves \cite{21,Geng,Branz}. While the vector component of $f_{0} (1710)$ might have the origin of the enhancement seen in $J/\psi\longrightarrow \gamma\varphi\omega$ near threshold observed at BES \cite{21,C. Garcı́a-Recio,M. Ablikim}. The scenario of these states will may be clear in future experiments like PANDA at FAIR. \\

Beside these, we have also calculated the decay width of remaining dimesonic states and listed in Table-IV, these states may be identified in the future experimental measurements.\\

\begin{table*}[]
\caption{Decay width of dimesonic (molecular) states(in MeV). The decay width is calculated for  experimentally seen decay modes listed in PDG \cite{21}. The other dimesonic states are also tabulated with isospin-spin (I,S).}
\begin{tabular}{c c c c c c c c }
\hline 
States & \multicolumn{7}{c}{Dimesonic decay widths (in MeV)}\tabularnewline
\cline{2-8}
 & \multicolumn{3}{c}{(in different decay modes) } & total & PDG \cite{21}  & others  & current  \tabularnewline 
&&&&width&& & status \tabularnewline
&&&&&& & in PDG \cite{21} \tabularnewline
\hline 
$f{}_{0}(980)_{(k\overline{k})}$ & $(49){}_{\pi\pi}$ & $(33){}_{\pi\eta}$ & $(12){}_{\pi\omega}$ & 94 &  50-100 & $91_{-20}^{+30}$  \cite{Ecklund}  & OK  \tabularnewline
&&&&&& 108 \cite{Rui Zhang}   &  \tabularnewline

$b{}_{1}(1235)_{(\eta\overline{\rho})}$ & $(89){}_{\pi\omega}$ & $(43){}_{\pi\phi}$ & - & 132 & $142\pm9$ & $151\pm31$ \cite{Fukui}   & OK  \tabularnewline
&&&&&& $113\pm12$ \cite{Weidenauer} \tabularnewline

$h{}_{1}(1380)_{(k\overline{k*})}$ & $(109){}_{\pi\pi}$ & $(71){}_{\pi\eta}$ & $(88){}_{\pi\omega}$ & 268 & $91\pm30$& $170\pm80$ \cite{43}  & NC! \tabularnewline
 
$a{}_{0}(1450)_{(\rho\overline{\omega})}$ & $(37){}_{\pi\eta}$ & $(32){}_{K\overline{K}}$ & $(25){}_{\pi\eta'}$ & 94 & $265\pm13$ & $265\pm30$ \cite{Amsler}    & OK \tabularnewline
&&&&&&$196\pm10$ \cite{Bugg} \tabularnewline

$f{}_{0}(1500)_{(\rho\overline{\rho})}$ & $(50){}_{\pi\pi}$ & $(29){}_{\eta\eta}$ & $(31){}_{K\overline{K}}$ & 110 & $109\pm7$ & $108\pm33$ \cite{24}   & OK \tabularnewline
&&&&&&$66\pm10$ \cite{Antinori} \tabularnewline

$f{}_{2}'(1525)_{(\rho\overline{\rho})}$ & $(25){}_{K\overline{K}}$ & $(23){}_{\eta\eta}$ & $(34){}_{\pi\pi}$ & 82 &  $76\pm10$ & $69\pm22$ \cite{Aguilar}  & OK \tabularnewline
&&&&&&$75\pm4$ \cite{Bai}\tabularnewline

$f{}_{2}(1565)_{(\omega\overline{\omega})}$ & $(33){}_{\pi\pi}$ & $(25){}_{K\overline{K}}$ & $(23){}_{\eta\eta}$ & 81 &  $134\pm8$ &$113\pm23$ \cite{Amsler}   & NC!   \tabularnewline
&&&&&&$119\pm24$ \cite{24} \tabularnewline

$h{}_{1}(1595)_{(\eta'\overline{\omega})}$ & $(63){}_{\omega\eta}$ & - & - & 63 &  $384\pm60$ & $384\pm60$ \cite{Eugenio}  & NC!\tabularnewline
 
$a{}_{2}(1700)_{(k*\overline{k*})}$ & $(75){}_{\pi\eta}$ & $(69){}_{K\overline{K}}$ & $(37){}_{\rho\omega}$ & 181 & $194\pm40$ & $151\pm22$ \cite{K. Abe}   & NC! \tabularnewline
&&&&&& $187\pm60$ \cite{Acciarri} \tabularnewline

$f{}_{0}(1710)_{(k*\overline{k*})}$ & $(45){}_{K\overline{K}}$ & $(43){}_{\eta\eta}$ & $(55){}_{\pi\pi}$ & 143 &  $135\pm7$ & $139_{-12}^{+11}$ \cite{Uehara}  & OK \tabularnewline
&&&&&&$145\pm8$ \cite{25} \tabularnewline

$f{}_{2}(1810)_{(\phi\overline{\omega})}$ & $(75){}_{\pi\pi}$ & $(62){}_{\eta\eta}$ & $(59){}_{K\overline{K}}$ & 196 &  $197\pm22$ & $228_{-20}^{+21}$ \cite{Uehara}   & NC! \tabularnewline

\hline 
${(\eta\overline{\eta})_{(0,0)}}$ & $(756){}_{\pi\pi}$ & $(554){}_{\pi\eta}$ & $(286){}_{K\overline{K}}$ & 1596 &   - & -  & - \tabularnewline

${(\eta\overline{\eta'})_{(0,0)}}$ & $(477){}_{\pi\pi}$ & $(411){}_{\pi\eta}$ & $(355){}_{K\overline{K}}$ & 1243 & - - &  - & - \tabularnewline

${(\eta'\overline{\eta'})_{(0,0)}}$ & $(841){}_{\pi\pi}$ & $(769){}_{\pi\eta}$ & $(714){}_{K\overline{K}}$ & 2324 &  -  &-   &  \tabularnewline

${(\eta\overline{\omega})_{(0,1)}}$ & $(114){}_{\pi\omega}$ & $(89){}_{\pi\eta}$ &  $(85){}_{\pi\rho}$ & 288 & - &   -  & - \tabularnewline

${(K\overline{K^{*}})_{(1,1)}}$ & $(74){}_{\pi\omega}$ & $(77){}_{\pi\rho}$ & $(74){}_{K\overline{K}}$ & 225 &  - & -  & - \tabularnewline

${(\eta\overline{\phi})_{(0,1)}}$ & $(70){}_{\pi\omega}$ & $(44){}_{\eta\omega}$ & $(70){}_{\pi\rho}$ & 184 & - &   -  & - \tabularnewline

${(\eta'\overline{\rho})_{(1,1)}}$ & $(87){}_{\pi\omega}$ & $(87){}_{\pi\rho}$ & $(58){}_{KK^{*}}$ & 232 & -&  -  & - \tabularnewline

${(\eta'\overline{\phi})_{(0,1)}}$ & $(175){}_{\pi\omega}$ & $(176){}_{\pi\rho}$ & $(139){}_{KK^{*}}$ & 490 &  - & -  & - \tabularnewline

${(\rho\overline{\rho})_{(0,1)}}$ & $(30){}_{\pi\rho}$ & $(30){}_{\pi\omega}$ & $(15){}_{KK^{*}}$ & 75 &  - & -  & - \tabularnewline

${(\rho\overline{\rho})_{(1,0)}}$ &  $(31){}_{\pi\rho}$  & $(15){}_{KK^{*}}$ & $(31){}_{\pi\omega}$ & 77 &   - & -  & - \tabularnewline

${(\rho\overline{\rho})_{(1,1)}}$ & $(29){}_{\pi\rho}$ & $(29){}_{\pi\omega}$ & $(14){}_{KK^{*}}$ & 72 &  - & -  & - \tabularnewline

${(\rho\overline{\rho})_{(1,2)}}$ &  $(35){}_{\pi\pi}$ & $(26){}_{K\overline{K}}$ & $(25){}_{\pi\omega}$ & 86 &   - & -  & - \tabularnewline

${(\omega\overline{\omega})_{(0,0)}}$ & $(46){}_{\pi\pi}$ & $(35){}_{K\overline{K}}$ & $(33){}_{\pi\omega}$ & 114 &  - & -  & - \tabularnewline

${(\omega\overline{\omega})_{(0,1)}}$ & $(30){}_{\pi\rho}$ & $(15){}_{KK^{*}}$ & $(29){}_{\pi\omega}$ & 74  &  - & -  & - \tabularnewline

${(K^{*}\overline{K^{*}})_{(0,1)}}$ & $(50){}_{\pi\omega}$ & $(40){}_{\eta\omega}$ & $(36){}_{KK^{*}}$ & 126  &  - & -  & - \tabularnewline

${(K^{*}\overline{K^{*}})_{(0,2)}}$ & $(60){}_{\pi\pi}$ & $(49){}_{K\overline{K}}$ & $(47){}_{\pi\omega}$ & 156 &  - & -  & - \tabularnewline

${(K^{*}\overline{K^{*}})_{(1,0)}}$ & $(52){}_{\pi\pi}$ & $(43){}_{K\overline{K}}$ & $(42){}_{\pi\omega}$ & 137  &  - & -  & - \tabularnewline

${(K^{*}\overline{K^{*}})_{(1,1)}}$ & $(48){}_{\pi\omega}$ & $(48){}_{\pi\rho}$ & $(35){}_{KK^{*}}$ & 131  &  - & -  & - \tabularnewline

${(\phi\overline{\phi})_{(0,0)}}$ & $(21){}_{\pi\pi}$ & $(18){}_{K\overline{K}}$ & $(9){}_{K^{*}K^{*}}$ & 48  &  - & -  & - \tabularnewline

${(\phi\overline{\phi})_{(0,1)}}$ & $(19){}_{\pi\omega}$ & $(20){}_{\pi\rho}$ & $(16){}_{KK^{*}}$ & 55  &  - & -  & - \tabularnewline

${(\phi\overline{\phi})_{(0,2)}}$ & $(28){}_{\pi\pi}$ & $(25){}_{K\overline{K}}$ & $(24){}_{\pi\omega}$ & 77 &  - & -  & - \tabularnewline

${(\rho\overline{\omega})_{(1,1)}}$ & $(29){}_{\pi\omega}$ & $(40){}_{\pi\pi}$ & $(14){}_{KK^{*}}$ & 83 &  - & -  & - \tabularnewline

${(\rho\overline{\omega})_{(1,2)}}$ & $(34){}_{\pi\pi}$ & $(26){}_{K\overline{K}}$ & $(25){}_{\pi\omega}$ & 85 &  - & -  & - \tabularnewline

${(\phi\overline{\rho})_{(1,0)}}$ & $(51){}_{\pi\pi}$ & $(42){}_{K\overline{K}}$ & $(41){}_{\pi\omega}$ & 134 &  - & -  & - \tabularnewline

${(\phi\overline{\rho})_{(1,1)}}$ & $(47){}_{\pi\omega}$ & $(47){}_{\pi\rho}$ & $(35){}_{KK^{*}}$ & 129  &  - & -  & - \tabularnewline

${(\phi\overline{\rho})_{(1,2)}}$ & $(81){}_{\pi\pi}$ & $(67){}_{K\overline{K}}$ & $(64){}_{\pi\omega}$ & 212 &  - & -  & - \tabularnewline

${(\phi\overline{\omega})_{(0,0)}}$ & $(53){}_{\pi\pi}$ & $(44){}_{K\overline{K}}$ & $(42){}_{\pi\omega}$ & 139 &  - & -  & - \tabularnewline

${(\phi\overline{\omega})_{(0,1)}}$ & $(39){}_{\pi\phi}$ & $(35){}_{KK^{*}}$ & $(47){}_{\pi\omega}$ & 121 & - & -  & - \tabularnewline

\hline
(NC!- Not Confirmed)
\end{tabular}

\end{table*}

\section{Conclusion}
In this paper, we are able to calculate the S-wave masses, digamma and decay widths of the dimesonic states in the light meson sector. Here, we would like to say that the dimesonic model becomes more accurate if binding energy of constituents is small compared to their masses. We compared many states which are experimentally  observed and predicted theoretically and having promising non- $q\overline{q}$ structure with dimesonic states. Calculated results are fairly in good agreement with experimental measurements as well as theoretical predictions. On the basis of mass spectra, decay width and in some cases digamma width, we are able to identify some dimesonic (meson-antimeson) states. We have calculated the decay widths for experimentally observed decay modes for each compared states.\\ 
In our present study, we strongly recommend the states $ f_{0}(980)$, $ b_{1}(1235)$, $ f_{2}'(1525)$, $ f_{0}(1710)$ as meson-antimeson molecules. Furthermore, the states $ h_{1}(1380)$, $ a_{0}(1450)$,  $ f_{2}(1565)$, $ h_{1}(1595)$ in calculation have been found with deviated masses and decay width from experimental measurements which rule out their candidature as dime-sonic molecules. Additionally, some states like  $ f_{0}(1500)$ $ a_{2}(1700)$ and $f_{2}(1810)$ calculated results have little bit variation in decay widths, but in some decay modes, decay width are fairly near to some experimental results \cite{43,Weidenauer,Acciarri}. Some states like $ a_{2}(1700)$, $f_{2}(1810)$ and $ f_{0}(1790)$ have unconfirmed experimental status, needing more attentions from experimentalist as well as form theoretician. \\

Finally, we have predicted the masses, decay widths and di-gamma widths of the S-wave dimesonic molecular states (light sector). Now we would like to employ this model to calculate the P-state masses and decay properties for light sector as well as for the heavy sector dimesonic systems.  \\		
\noindent {\bf Acknowledgements}

A. K. Rai would like to thanks Prof. Atsushi Hosaka from Osaka University for the useful discussion, and also acknowledge the financial support extended by D.S.T., Gov. of India  under SERB fast track scheme SR/FTP /PS-152/2012. \\

\end{document}